\documentclass[useAMS,usenatbib]{mn2e}
\usepackage{epsfig}   
\input{psfig}

\def\<#1>{\langle\hbox{#1}\rangle}
\def\etal{{et~al.}}
\def\kms{km~s$^{-1}$}

\def\eg{{e.g.}}

\def\simgt{\hbox{\rlap{\raise 0.425ex\hbox{$>$}}\lower 0.65ex\hbox{$\sim$}}}
\def\simlt{\hbox{\rlap{\raise 0.425ex\hbox{$<$}}\lower 0.65ex\hbox{$\sim$}}}

\def\deriv(#1/#2){\mathchoice{d#1\over d#2}
    {d#1/d#2} {d#1/d#2} {d#1/d#2}}
\def\pderiv(#1/#2){\mathchoice{\partial#1\over\partial#2}
    {\partial#1/\partial#2} {\partial#1/\partial#2} {\partial#1/\partial#2}}
\def\pderivp(#1/#2){\left(\pderiv(#1/#2)\right)}

\def\tderiv(#1/#2){\mathchoice{d#1\over d#2}{d#1/d#2}
                   {d#1/d#2}{d#1/d#2}}
\def\tderivp(#1/#2){\left(\tderiv(#1/#2)\right)}

\title[Star formation along Pisces-Cetus filaments]{Star formation
in galaxies along the Pisces-Cetus Supercluster filaments}
\author[Porter and Raychaudhury]{Scott C. Porter and
        Somak Raychaudhury\thanks{E-mail: scp@star.sr.bham.ac.uk;
              somak@star.sr.bham.ac.uk}\\
School of Physics and Astronomy, University of Birmingham, 
Birmingham B15~2TT, UK}

\begin{document}

\date{Accepted 2006 December. Received in original form 2006 August}

\pagerange{\pageref{firstpage}--\pageref{lastpage}} \pubyear{2006}

\maketitle

\label{firstpage}

\begin{abstract}
We investigate the variation of current star formation in galaxies as
a function of distance along three supercluster filaments, each
joining pairs of rich clusters, in the Pisces-Cetus supercluster,
which is part of the two-degree Field Galaxy Redshift Survey
(2dFGRS). We find that even though there is a steady decline in the
rate of star formation, as well as in the fraction of star forming
galaxies, as one approaches the core of a cluster at an extremity of
such a filament, there is an increased activity of star formation in a
narrow distance range between 3-4 $h_{70}^{-1}$~Mpc, which is 1.5--2
times the virial radius of the clusters involved.  This peak in star
formation is seen to be entirely due to the dwarf galaxies ($-20\!<\!
M_B \!\le\! -17.5$).  The position of the peak does not seem to depend
on the velocity dispersion of the nearest cluster, undermining the
importance of the gravitational effect of the clusters involved. We
find that this enhancement in star formation occurs at the same place
for galaxies which belong to groups within these filaments, while
group members elsewhere in the 2dFGRS do not show this effect.  We
conclude that the most likely mechanism for this enhanced star
formation is galaxy-galaxy harassment, in the crowded infall region of
rich clusters at the extremities of filaments, which induces a burst
of star formation in galaxies, before they have been stripped of their
gas in the denser cores of clusters. The effects of strangulation in
the cores of clusters, as well as excess star formation in the infall
regions along the filaments, are more pronounced in dwarfs since they
more vulnerable to the effects of strangulation and harassment than
giant galaxies.
   
\end{abstract}



\section{Introduction}

Star formation within individual galaxies is triggered by
gravitational instabilities, induced by pressure variations due to
density waves or shear forces, random cloud collisions or shocks due
to stellar winds and supernova explosions \citep[\eg][]{ken1998}. 
It is
becoming increasingly clear that, in addition to these, the influence
of external effects, related to the immediate environment of the
galaxy, are equally important in inducing star formation. Among the
latter are the effects of merger and collision, stripping, accretion
and harassment, which are responsible for both the triggering of star
formation and quenching the process as the galaxy loses its gas
content.

It has been shown that the average rate of star formation in galaxies
varies widely with redshift \citep[\eg][]{madau98,pog06}, and so does
the relative morphological fraction of galaxies
\citep[\eg][]{abr96,goto-bo03}, providing direct evidence of the
chemical evolution of galaxies with time.  At any given redshift,
however, the rate of star formation in a galaxy, and the morphological
content of the galaxy population, change with the local environment.
Early studies showed that the relative morphological content of
galaxies in a given volume of space depends on the local projected
galaxy density \citep[\eg][]{dressler1980}.  Even though there is a
clear link between the average star formation rate (SFR) of a
galaxy and its morphology \citep[\eg][]{menan06}, detailed
photometric and spectroscopic studies covering a large range of
redshift \citep[\eg][]{ravi06,yuan05} show that the they cannot be used as
proxies of one another. 

Effects of this nature have been extensively studied in galaxy
clusters, where there is clear evidence of a systematic variation of
properties like morphology, shape and SFR with distance from the
centre of the cluster \citep[\eg][]{ms77,whit93,deprop2003}, even at
higher redshifts \citep[\eg][]{gerken04}.  Clearly there are various
factors to consider in quantifying both local and global environments,
and galaxy evolution, in terms of parameters that reflect the ages of
stellar populations and recent history of activity, as well as global
morphology and the presence of small-scale structure. Untangling the
effects of these various factors require large samples, and the work
in this field has only just begun
\citep[\eg][]{chris2005}.

\begin{figure*}
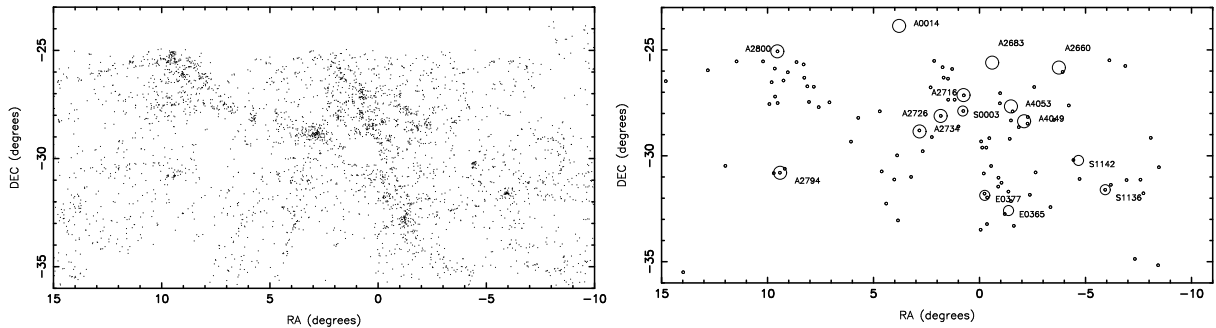

\begin{center}
\psfig{figure=scp-pisces-fig1a.ps,angle=-90,width=0.45\hsize}
\psfig{figure=scp-pisces-fig1b.ps,angle=-90,width=0.45\hsize}
\end{center}
\medskip
 \caption{ (a, Left) All 2dFGRS Galaxies within 1840 kms$^{-1}$ ($3
 \sigma$) of the mean recessional velocity of the Pisces-Cetus
 supercluster. The filamentary structure is clearly visible.  (b,
 Right) Clusters of galaxies belonging to Pisces-Cetus supercluster
 filaments within the 2dFGRS region and within 1840 kms$^{-1}$ ($3
 \sigma$) of the mean supercluster velocity.  The largest circles
 represent Abell clusters, and the medium-sized ones are supplementary
 Abell and Edinburgh-Durham clusters within the same redshift range.
 The groups of galaxies \citep[][2PIGG]{eke2004} within the same
 velocity bounds, identified from a friends-of-friends analysis of the
 2dFGRS are shown as the smallest circles.
\label{gals}
}
\end{figure*}

It is therefore interesting to investigate further into the nature of
the evolution of galaxies, and which aspects of their physics respond
to the local environment of the galaxy, and in what way. In 
the cores of rich clusters,
for instance, it has been long known that 
the star formation rate in galaxies
is highly suppressed
\citep{dressler1980,whit93}, and the current rate of star
formation progressively decreases as the environment becomes denser
\citep[\eg][]{balogh02,kauf2004}.  There seems to be a break in the
relation between local projected density and SFR at a projected
density about 1~Mpc$^{-2}$, and out to 3--4 times the
virial radius of the cluster, the SFR remains well below that of the
field \citep{couch1987,balogh98,lewis02,gomez03}. This effect seems to
be more pronounced among giant galaxies ($M_R\!<\-20$), whereas dwarfs
are found to be passive only within the virial radius of the cluster,
outside which almost all dwarfs are found to be star-forming
\citep{haines06apj}.

Given that only a minority of galaxies lie in clusters, is this a
representative picture of galaxy evolution? The projected local
density of galaxies seems to be a na\"{\i}ve representation of the environment
of the galaxy, given the complex histories of galaxies.  Observations
and simulations of the large-scale structure of the Universe have also
shown galaxies to be arranged in a network of filaments and voids
\citep{einasto94,virgo01}. In simulations,
galaxies seem to form in groups on this filamentary network, and with
time the groups merge with others as they fall into rich clusters
which lie at the intersections of these filaments. Indeed,
observational evidence of such infall of groups, preferentially from
the directions to neighbouring rich clusters, have been found in
nearby systems
\citep[\eg\, the Coma cluster][]{adami05}. 
Much of the evolution of the galaxies thus happens in the
group environment, but the potential effect of the larger-scale
environment is undeniable, given that it would fundamentally affect
the history and dynamics of the parent group of the galaxy.

In this paper we investigate the nature of star formation in galaxies
along supercluster filaments, paying attention to the whether the
galaxy is a dwarf or a giant, 
and the nature of the group or cluster the galaxy
belongs to, as well as distance from its nearest rich cluster. In
\S\ref{sec:sc} we identify the galaxies that belong to the highly
filamentary structures of the Pisces-Cetus supercluster
\citep{porter2005}, and classify them according to their membership in
groups. \S\ref{sec:sfr1} presents the dependence of star formation
properties (and also the fraction of passive galaxies) as a function
of distance from the nearest cluster, for all galaxies, as well as
subsets according to luminosity, group membership and the richness of
the nearest cluster. The implications of these in terms of the insight
they offer for the dependence of star formation on environment are discussed
in \S\ref{sec:dis},
and broad conclusions are drawn in \S\ref{sec:conc}. 

The cosmology used is $\Omega_M=1$ and $H_0\!=\!70$~km~s$^{-1}$
Mpc$^{-1}$, though the trends we present do not depend on
these parameters.

\section{The Pisces-Cetus Supercluster filaments}
\label{sec:sc}

A supercluster filament is a snapshot of a cluster in formation, and
the study of the properties of galaxies as a function of position
along a filament is expected to provide insights into the processes
involved in galactic evolution and cluster formation. Here we have
chosen an ensemble of filaments, lying within the volume of the
Pisces-Cetus supercluster, covered by the two-degree Field Galaxy
Redshift Survey
\citep[][2dFGRS]{2dFGRS-final}. 

The clusters comprising the Pisces-Cetus supercluster within the
2dFGRS region, taken from the minimal spanning tree (MST) analysis of
\citet{br00,somak07}, form the core of our sample.  A complete list of
positions and redshifts of the entire Pisces-Cetus supercluster, which
extends some 15 degrees to the north of the 2dFGRS region, can be
found in \citet{porter2005}.  The filamentary nature of the 2dFGRS
region can clearly be seen in the distribution of galaxies in the
groups in Fig.~\ref{gals}a, where the 2dFGRS galaxies within 1840
\kms\ (three times the velocity dispersion of the clusters belonging
to the supercluster), are plotted. The 2PIGG groups found within the
same velocity bounds
\citep{eke2004} also show the same large-scale structure.
The rich clusters, plotted as large circles on the same plot, can be
seen to be mostly along these filaments forming the supercluster
structure.

\subsection{Filament membership}
\label{sec:mem}

The 2dFGRS Percolation-Inferred Galaxy Group (2PIGG) catalogue
\citep{eke2004} is a list of galaxy groups and their members, based on
the North and South strips of the 2dFGRS redshift catalogue. It
consists of approximately 190,000 galaxies, about 55$\%$ of which are
members of the $\sim$29,000 galaxy groups and clusters found using a
friend-of-friends algorithm. This work leaves out about 20\% of
2dFGRS galaxies in under-sampled regions.  In this paper, we use this
subset of 2dFGRS galaxies, and hereafter, 2dFGRS refers to this
subsample. Due to the varying survey limit,
the absolute magnitude limit of the 2dFGRS
at the mean redshift of the Pisces-Cetus supercluster varies across
the supercluster region between $B_J\!=\! -17.2$ to $-17.9$.

From this sample, galaxies which belong to the three most prominent
filaments longer than 20 $h_{70}^{-1}$~Mpc, seen in Fig.~\ref{gals},
namely those joining Abell~2800--Abell~2734, Abell~2734--EDCC~0365 and
EDCC~0365--Abell~2716 (clusters from \citealt{aco89,edcc92}), were
extracted (hereafter we refer to the clusters only with the 'A' and
'E' prefixes). The filamentary structure of the supercluster is
further discussed in \citet{porter2005}.

\begin{figure}
\begin{center}
\psfig{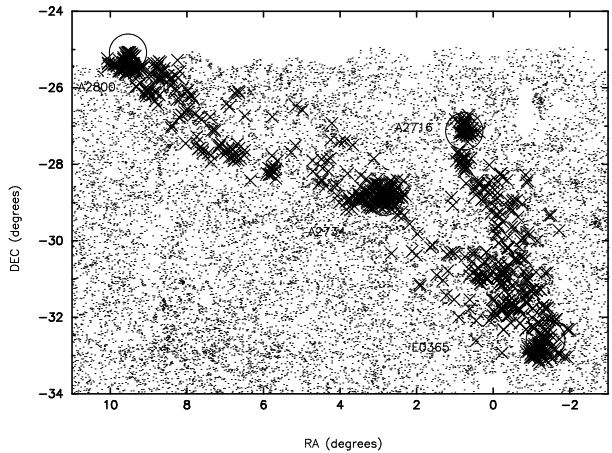}
\end{center}
\medskip
 \caption{The galaxies that are members of
  the three filaments of the Pisces-Cetus supercluster,
 as defined in the
 text (\S\ref{sec:mem}), 
 are shown as crosses. All galaxies within the 2dFGRS in the
 region are shown as dots.}
\label{fillsky}
\end{figure}

We define the extent of each filament as a prolate spheroid, with the
centres of the two clusters involved being at each end, the distance
between them being the major axis of the spheroid, and
$6\,h_{70}^{-1}$~Mpc being the semi-minor axis.  All galaxies falling
inside this spheroid were taken to be members of the filament.  So as
to not omit cluster galaxies at each end of the filament, we added the
galaxies found from a percolation analysis to belong to the
corresponding 2PIGG ``group'' in \citet{eke2004}.  The resulting 965
filament members can be seen in Fig.~\ref{fillsky}, highlighted as crosses 
in the general field of all 2dFGRS galaxies in this part of the sky.

The distance between two clusters, and between a galaxy and a cluster,
was calculated as the comoving proper distance between them
\citep[see][]{hogg99}. We used the measured redshift of each
galaxy as a measure of its distance from us, except if it is a member
of a rich cluster, where the mean redshift of the cluster was used.

\section{Star formation properties of galaxies along Supercluster filaments}
\label{sec:sfr1}

Having identified the major filaments in the Pisces-Cetus supercluster, 
and the galaxies, groups and clusters that lie on them, in this section
we look at their rate of star formation (represented by the $\eta$ parameter)
as a function of distance 
along the filaments.

\begin{figure}
\begin{center}
\psfig{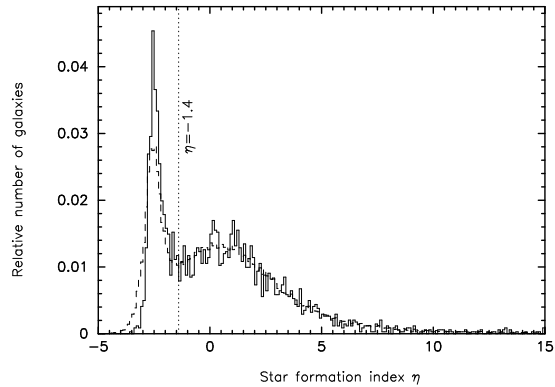}
\end{center}
\medskip
 \caption{A histogram of the $\eta$ parameter for all 2dFGRS galaxies
 within $z<0.1$ is shown as the dashed line. Of these, the solid
 histogram shows galaxies within the bounds of the Pisces-Cetus
 supercluster and 1840 km~s$^{-1}$ ($3\sigma$) of the mean recessional
 velocity of the supercluster.}
\label{hist2}
\end{figure}

\subsection{The $\eta$ parameter}
\label{sec:eta}

Since our filaments are in the region covered by the 2dFGRS, we opt to
use a simple parameter, related to the star formation rate, that has
been derived for most 2dFGRS galaxies.
\citet{madgwick2002} used Principal Component Analysis
of de-redshifted 2dF galaxy spectra to define a parameter $\eta$, a
linear combination of the first two principal components, which
correlates well with the equivalent width of the H$\alpha$
[EW(H$\alpha$)] emission line, which in turn is a measure of SFR
\citep[e.g.,][]{ken1983,gall1984,moustakas2006}. With some scatter,
$\eta\approx -2.0$ correspond to no H$\alpha$ emission at all,
increasing to 
$\eta\approx 7$
for EW(H$\alpha$) $=50$\AA\ \citep{madgwick2003}.

A histogram of the $\eta$ parameter for all 2dFGRS galaxies
(Fig.~\ref{hist2}, dashed line) shows two distinct peaks in the
distribution, at $\eta\!\sim \!-2.5$ and at $\eta\!\sim\! 0.05$. This
reflects the well-known bimodality of the local galaxy population, of
red, passive galaxies, and bluer, star forming galaxies
\citep[\eg][]{kauf2004,balogh04}.  The dip or divide between the two
peaks is at $\eta\!\sim\! -1.4$. The first peak only contains
$\sim$30$\%$ of the galaxies with $z\!<\! 0.1$.
\citet{madgwick2003} show that the first narrow peak 
corresponds to E/S0 galaxies while the second broader peak represents
later-type galaxies. Indeed, they also find a tight correlation
between $\eta$ and the birth rate parameter $b$, which is the ratio of
current to past-averaged SFR.  The dividing line between the two peaks
at $\eta\!\sim\! -1.4$ corresponds to $b\!\sim\! 0.1$, the value where
the current SFR is only a tenth of that in the past.
The $\mu_\ast$ parameter of \citet{lewis02}, which is the SFR
normalised to the characteristic Schechter luminosity $L_\ast$, is
shown to be $\mu_\ast\!=\!0.087$~EW(${H\alpha}$).  Therefore,
$\eta\!\sim\! -1.4$ would correspond to $\mu_\ast\sim 0.4$.

Also shown in Fig.~\ref{hist2} is the corresponding histogram (solid
line) for the galaxies within the supercluster region as defined in
\S\ref{sec:sc}.  
The supercluster histogram has a higher peak
corresponding to passive galaxies.  This would be consistent with a
higher proportion of early type galaxies in the supercluster
filaments, according to the morphology-density relation, in accord
with the SDSS-derived results of \citet{tan2004}.

In this study of the star formation properties of galaxies, we look at the
variation of two quantities as a function of position along the filaments
and distance away from the nearest cluster:
\begin{enumerate}
\item the value of the $\eta$ parameter, which correlates with the 
current SFR of the galaxy, and
\item the fraction of galaxies with $\eta < -1.4$, 
which represents the fraction of non-star-forming (passive) galaxies
in a subsample.
\end{enumerate}

\subsection{Star formation along the A2734-A2800 filament}

We begin by looking in detail at the most prominent filament in the
Pisces-Cetus supercluster, that linking A2734 and A2800, which is
about 32 $h_{70}^{-1}$~Mpc long.  The comoving distance of each of the
filament galaxies from one end (A2734) was calculated as described
above.  Members of A2800 were assigned a distance equal to the
separation of the two clusters minus their projected distance from the
centre of A2800.

Along the filament, we chose bins of varying size according to the
numbers of galaxies available. Closer to the two clusters, we chose
smaller bins of 0.3--0.6 $h_{70}^{-1}$~Mpc to sample the rapid
variation of SFR within the virial radius of the cluster. For each
bin,  the mean distance of all member galaxies, and the mean
$\eta$, were calculated.

\begin{figure}
\begin{center}
\psfig{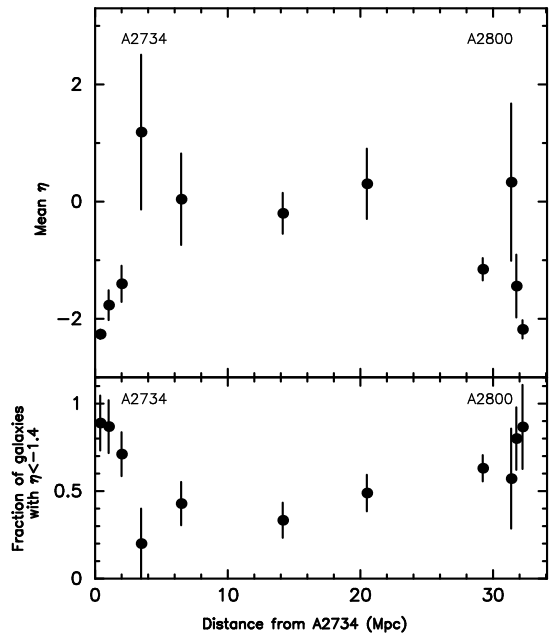}
\end{center}
\medskip
 \caption{For the filament of galaxies linking A2734 and A2800, the
 mean $\eta$ is plotted as a function of the distance from the centre
 of A2734.  The ``field'' value, the mean $\eta$ for all galaxies of
 the 2dFGRS, is around zero.  The bottom panel shows the fraction of
 passive galaxies ($\eta\! < \! -1.4$) in each bin.  It can be seen,
 as expected, that the SFR is lowest at the cores of the two rich
 clusters, where the number of passive galaxies is also the highest. 
 In addition, the value of 
 mean $\eta$ seems to peak between 3--4 $h_{70}^{-1}$~Mpc from each
 end.}
\label{single}
\end{figure}  

Fig.~\ref{single} shows the resulting plot of mean $\eta$ as a
function of distance from the centre of A2734. We also plot, in the
bottom panel, the fraction of passive ($\eta\! < \! -1.4$) galaxies in
each bin.  It can be seen, as expected, that the SFR is minimum at the
cores of the two rich clusters, where the number of passive galaxies
is the highest. The mean value of the $\eta$ parameter increases with
distance to approximately the field value, since the mean $\eta$ for
all galaxies of the 2dFGRS is around zero.

However, it can also be seen that between 3--4 $h_{70}^{-1}$~Mpc from
the centre of A2734, there is a peak in the value of mean
$\eta$. Correspondingly, there appears to be another peak at the other
end of the plot, at a slightly closer distance from the centre of
A2800, although, within errors, it does not rise above the field. To
investigate the reality of the signal, we will now stack all three
filaments in our sample, and fold Fig.~\ref{single} such that we
consider distances along the filament from the nearest cluster.

\begin{figure*}
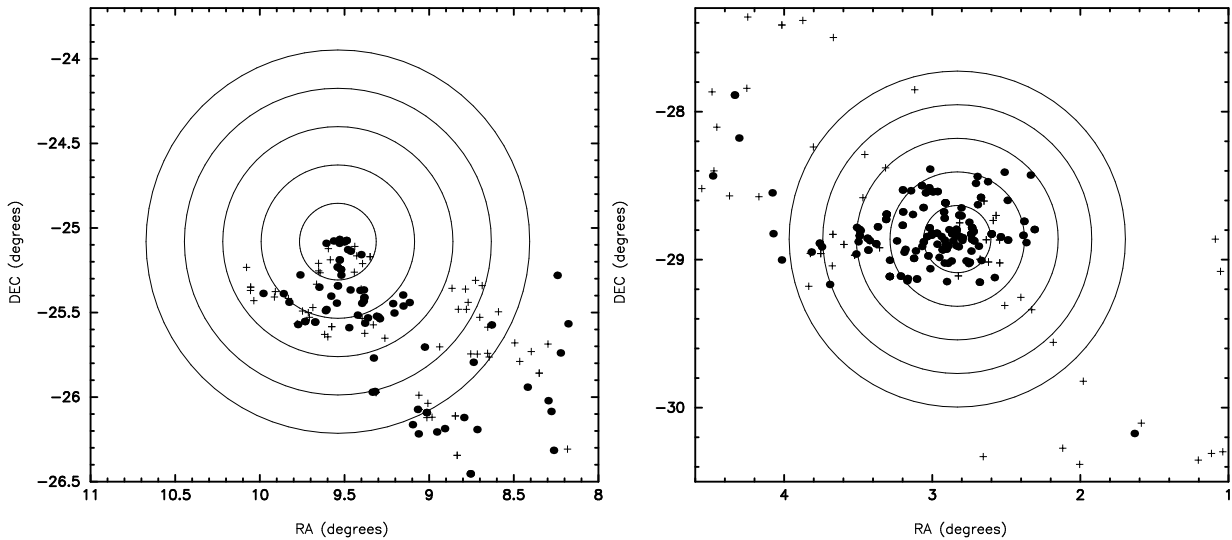

\begin{center}
\psfig{figure=scp-pisces-fig5a.ps,width=0.4\hsize,angle=-90}
\psfig{figure=scp-pisces-fig5b.ps,width=0.4\hsize,angle=-90}
\end{center}
\medskip
\caption{
The distribution of passive ($\eta\!<\! -1.4$) galaxies shown as
filled circles and star forming ($\eta\!>\! -1.4$) galaxies shown as
crosses in the region surrounding (left) A2800 and (right) A2734. Only
galaxies belonging to the A2734--A2800 filament are shown. The
concentric circles represent $1\, h_{70}^{-1}$~Mpc intervals.
\label{zoom1}
}
\end{figure*}

Fig.~\ref{zoom1} shows the distribution of the
passive and star forming galaxies, in the immediate vicinity
of the clusters A2800 and A2734
respectively. The concentric circles are at $1\, h_{70}^{-1}$~Mpc
intervals. It can be seen that in the 3--4$\, h_{70}^{-1}$~Mpc
annulus there are more star forming than passive galaxies for both
clusters.

\begin{table}
\begin{center}
\caption{Virial radii for Pisces-Cetus clusters involved in this analysis
\label{tab:vir}
}
\footnotesize
\medskip
\begin{tabular}{lccc}
\hline
Cluster & $\sigma_r$    & $N_{\rm gal}$  &Virial radius    \\
        & ({\kms})  &  &($h_{70}^{-1}$~Mpc)            \\
\hline
A2734   &848  &141                 &2.5
\\
A2716   &812  &77                  &2.4
\\
A2800   &567  &79                  &1.7
\\
E0365   &442  &78                  &1.3
\\
\hline
\end{tabular}
\end{center}
\noindent
{\footnotesize}
\end{table}

\subsection{Star formation 
properties of galaxies in all three filaments combined}

\begin{figure}
\begin{center}
\psfig{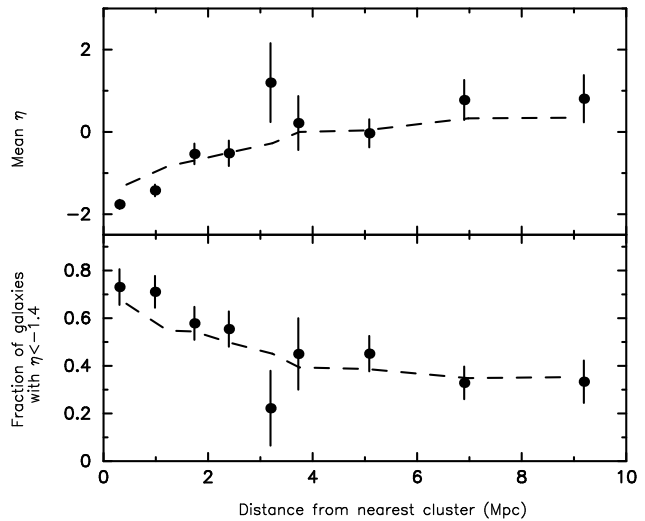}
\end{center}
\medskip
 \caption{ (a, Top) For the three filaments combined, the mean $\eta$
 as function of distance from the nearest cluster is shown.  The
 dashed line shows the mean $\eta$ as function of distance from the
 nearest 2PIGG group with $\ge 30$ members for all 2dFGRS
 galaxies.  (b, Bottom) For the same galaxies as above, the fraction of
 these galaxies with an $\eta\!<\!-1.4$ 
is shown as a function of distance
 from the nearest cluster.  The dashed line showing the same fraction
 as a function of distance from the nearest 2PIGG group with 
 $\ge 30$  members for the whole 2dFGRS.
\label{peak-filament}
}
\end{figure}

Here we combine the three filaments longer than 20 $h_{70}^{-1}$~Mpc,
namely those joining A2800--A2734, A2734--E0365 and E0365--A2716, as
seen in Fig.~\ref{fillsky}.  All galaxies belonging to the three
filaments were grouped in bins of varying size according to their
distance from the nearest cluster (representing the extremities of the
filaments). As above, we plot the mean value of $\eta$ and the
fraction of passive galaxies in each bin in
Fig.~\ref{peak-filament}.  For comparison between the trend of the
SFR with distance for the filament galaxies, with that of galaxies
elsewhere, we compute ``field'' values from the whole 2dFGRS, where
distances are calculated from the nearest 2PIGG group of $\ge 30$
members (the equivalent of a rich cluster), shown as the dashed line.

While the ``field'' values of mean $\eta$ increase steadily with
distance from the centre of the nearest cluster, the galaxies in the
filaments, binned in the same way, show a sharp peak between 3--4
$h_{70}^{-1}$~Mpc, 
confirming our finding from the single filament
above. 

The virial radius of a cluster can be approximately derived from its
radial velocity dispersion as $R_{\rm vir} \sim 0.002\, \sigma_r
\,h^{-1}$ Mpc \citep{girardi98}, where $\sigma_r$ is given in {\kms}.
For the four clusters that lie at the intersection of the three
filaments involved in this study, we computed the virial radii from
this relation, which are  given in Table~\ref{tab:vir}. 
The radial velocity dispersions were calculated using the algorithm 
of \cite{danese80}, from 2dFGRS redshifts.
The mean virial radius
for these clusters is about 2~$h_{70}^{-1}$~Mpc,
which means that the abrupt peak of enhanced star
formation appears at about 1.5--2 virial radii from the centres of the
clusters concerned.

A corresponding dip in the fraction of passive
galaxies is also seen at this distance. Of course, star formation is
suppressed as the galaxy approaches the core of the cluster, leading
to a low mean $\eta$ and a higher fraction of passive galaxies
interior to this distance.  The implications of these trends are
discussed in \S\ref{sec:dis}.

\subsection{Star formation in giant and dwarf galaxies}

As samples of galaxies with estimated star formation properties grow
larger, it would become easier to study the relation between star
formation in a galaxy and various properties of its environment, and
those of the galaxies themselves. From SDSS-DR4, for example,
\citet[\eg][]{haines06apj} find that among giant ($M_R\!<\! -20$)
galaxies, the fraction of passive galaxies ($H_\alpha$~EW$<$4\AA)
declines steadily out to 3--4 times the virial radius $R_V$, whereas
the fainter dwarfs are passive only within $R_V$, outside of which
they are overwhelmingly star-forming.  The fact that passive dwarf
galaxies are rare even in the infall regions of rich clusters (where
low velocities would make mergers more probable) would imply either,
that star formation in dwarfs is not easily quenched by mergers, or
more plausibly, that the mergers between dwarfs are not very
common. Instead, dwarfs seem to be more vulnerable, in the more
crowded regions, of the stripping away of the outlying diffuse gas,
which is the potential fuel for star formation, by the more massive
haloes of the giants they become gravitationally bound to
\citep[e.g.,][also known as {\it strangulation}]{larson80}.

\begin{figure}
\begin{center}
\psfig{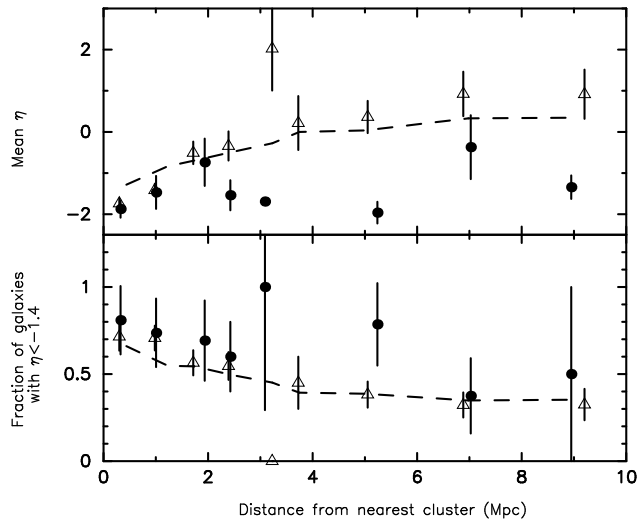}
\end{center}
\medskip
 \caption{ Star formation in giant galaxies
($M_B \!\le\! -20$), compared to that in dwarfs
($-20\!<\! M_B \!\le\! -17.5$), in the sample used in
Fig.~\ref{peak-filament}:
(a, Top) The mean star formation rate parameter $\eta$ is shown
 as function of distance from the nearest cluster, for 
dwarfs (open triangles) and giants (filled circles). It appears that the
enhanced star formation at $\sim$3 Mpc appears only in the dwarfs.
 (b, Bottom) The fraction of
 the galaxies with  $\eta\!<\! -1.4$ (passive galaxies) 
 is shown as a function of distance
 from the nearest cluster.
 There are no passive dwarf galaxies
 in the bin containing the peak in mean $\eta$.  
\label{fil-mags}
}
\end{figure}

Here, we ask whether this trend in the SFR of galaxies along the
filament depends on whether the galaxy in question is a giant or a
dwarf.  We divide the galaxies plotted above in
Fig.~\ref{peak-filament} into two subsamples: giant galaxies ($M_B
\!\le\! -20$) and dwarf galaxies ($-20\!<\! M_B \!\le\!  -17.5$). As a
result, there are 119 giants and 846 dwarfs in the three filaments
used above. We plot the mean value of $\eta$ as a measure of star
formation, and the fraction of passive galaxies, as a function of
their distance from the nearest cluster, in Fig.~\ref{fil-mags}, with
dwarfs plotted as open triangles and giants as filled circles.

It seems that the peak in mean $\eta$ seen in
Fig.~\ref{peak-filament}, between 3--4 $h_{70}^{-1}$~Mpc, is 
almost entirely due to enhanced
star formation in dwarfs-- indeed there are no passive dwarf galaxies
in the bin containing the peak in mean $\eta$ (see bottom panel
of Fig.~\ref{fil-mags}. 

The more luminous galaxies ($M_B \!\le\! -20$) seem to have the same
deficiency in star formation at the very cores of clusters, and have
an overall lower rate of star formation than the dwarf galaxies, thus
not contributing much to the trend of $\eta$ with distance.  Although,
the mean $\eta$ with distance from the centre of the nearest cluster
of just these giant galaxies, seems to show a peak in activity in the
range 1.5--2 $h_{70}^{-1}$~Mpc, the small number of galaxies in that
bin ($N\!=\! 13$) put a large element of doubt on its validity.  These
results are discussed in more detail in \S\ref{sec:dis}.

\subsection{Enhanced star formation as a function of cluster
richness}
\label{sec:sfvd}

If the tidal influence of the galaxy cluster at the end of the filament
were responsible for the enhanced star formation seen above, then
we would expect to see a correlation of the position of the peak SFR
with the richness of the cluster.
One would indeed expect galaxies falling into clusters
of larger velocity dispersion to encounter tidal forces at larger
distances.

\begin{figure}
\begin{center}
\psfig{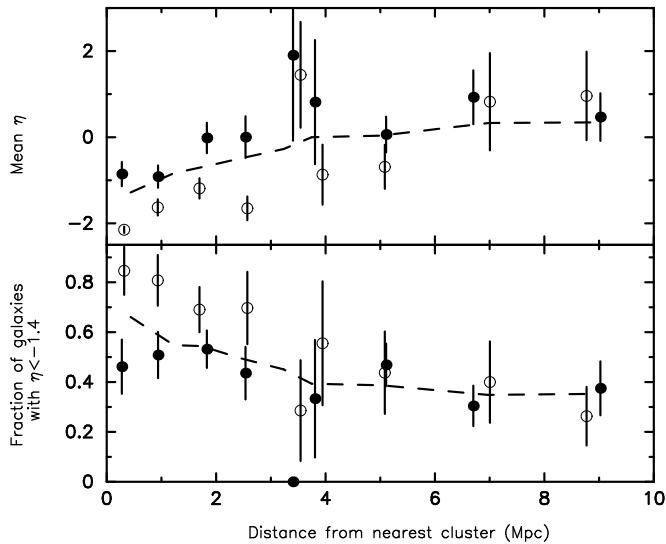}
\end{center}
\medskip
 \caption{The dependence of the mean star formation rate of galaxies,
 on the richness of clusters closest to them, is shown here, for the
 sample of three filaments used in Fig.~\ref{peak-filament}. (a, Top)
 Mean $\eta$ as a function of distance from the nearest cluster is
 shown, where galaxies closest to the low velocity dispersion clusters
 (A2800 and E0365) are shown as filled circles, while galaxies closest
 to the high velocity dispersion clusters (A2734 and A2716) are shown
 as open circles.  (b, Bottom) For the same galaxies as above, the
 fraction of the galaxies with $\eta\!<\!-1.4$ (passive galaxies) is
 shown as a function of distance from the nearest cluster.  The dashed
 line in both cases shows the corresponding parameter for all galaxies
 in the 2dFGRS, where the nearest ``cluster'' is defined as the
 nearest 2PIGG group with $\ge 30$ members.
\label{sfveldis}}
\end{figure}

We therefore split the four clusters that form the extremities of our
filaments into those with high velocity dispersions (A2716:
812~km~s$^{-1}$, A2734: 848~km~s$^{-1}$) and those with low velocity
dispersions (A2800: 567~km~s$^{-1}$, E0365: 426~km~s$^{-1}$).
Fig.~\ref{sfveldis} shows the resulting values for mean $\eta$ as a
function of distance for the high and low velocity dispersion cluster
samples. It can be seen that there is no significant difference in the
position of the low velocity dispersion cluster peak in mean $\eta$
(filled circles) and the peak in the high velocity dispersion cluster
peak (open circles). 

\subsection{Star formation in groups on inter-cluster filaments}

The suppression of star formation with increasing local density of
galaxies has been observed to occur in giant galaxies at very low
values ($\sim$1~Mpc$^{-2}$) of the local projected density
\citep[\eg][]{lewis02,gomez03},
which are typical of densities well outside the virialised regions in
clusters, indicating that the cause does not lie entirely in the
influence of the cluster environment. One suggestion is that much of
the evolution of galaxies occurs in groups during their life on the
filaments, before the groups are assimilated in clusters. Evidence
supporting this is found in the existence of galaxies with hot X-ray
emitting haloes in groups \citep{ewan01}.  If so, the trend seen in
the previous section would be different for filament galaxies that
belong to groups and those that are relatively isolated.

We could have looked for differences in star-forming properties
between galaxies belonging to groups and those not in groups on these
filaments, but the test proved inconclusive due to small numbers,
since only 8\% of galaxies in our sample do not belong to any 2PIGG
group of 4 or more members.

\begin{figure}
\begin{center}
\psfig{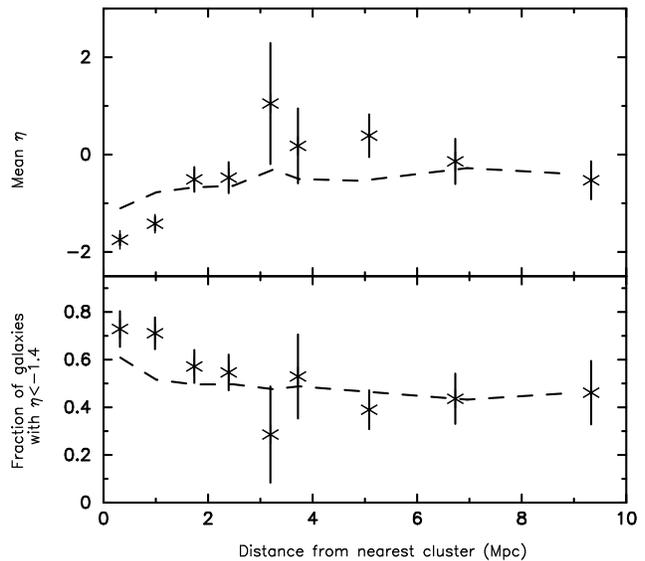}
\end{center}
 \caption{ Star formation in galaxies in 2PIGG groups: those in the
 Pisces-Cetus filaments compared to those in the whole of the 2dFGRS:
 (a, Top) The mean value of $\eta$ as a function of distance from the
 nearest cluster.  Filament galaxies which are members of groups (2PIGG
 groups, $N\ge 4$) are shown as crosses, while all galaxies in the
 2dFGRS which are members of similar 2PIGG groups are represented by
 the dashed line. The peak in $\eta$ is not seen in the latter
 category.  (b, Bottom) The fraction of passive filament galaxies
 ($\eta\!<\!-1.4$) as a function of distance from nearest cluster for
 the same samples.
\label{fil-groups}
}
\end{figure}

Instead, Fig.~\ref{fil-groups} compares the same quantities considered
above, for galaxies that belong to groups in these three filaments, to
those for similar group galaxies elsewhere in the 2dFGRS.  The dashed
line represents the variation of mean $\eta$ and passive galaxy
fraction as a function of distance from the nearest cluster for all
galaxies that belong to a 2PIGG group of four or more members (a
cluster being defined as a 2PIGG group of 30 or more members). The
crosses represent the galaxies belonging to our sample of three
Pisces-Cetus filaments.

The plots suggest that the star formation rate is more or less uniform
in galaxies which are members of groups in the entire 2dFGRS
catalogue, though at a slightly lower mean level than found in the
field (shown as the dashed line in Fig.~\ref{peak-filament},
irrespective of their distance from the nearest cluster, except for
those group galaxies within $\sim 3\,h_{70}^{-1}$~Mpc  of the centres of rich
clusters. In sharp contrast, the trend we find of a peak in SFR
between $3\! - \!4\,h_{70}^{-1}$~Mpc 
is seen in the group galaxies which
belong to filaments, indicating that the presence of the filaments
markedly influences star formation properties in galaxies that belong
to groups.  The variation is even more highlighted in the bottom
panel, showing the fraction of passive galaxies in each bin. The physical
processes that govern the excess of star formation in these galaxies clearly
are more related to the supercluster environment than the group environment.

\section{Discussion}
\label{sec:dis}

The most striking of our results are those seen in
Fig.~\ref{peak-filament}, where we look at galaxies which are members
of filaments connecting pairs of rich clusters of galaxies.  Firstly,
we see, as expected, the decline in SFR in these galaxies, from those
in the filament environment on the periphery of a cluster, to the
galaxies in the cluster core. This indicates that a physical mechanism
is at work that quenches star formation progressively as a galaxy
approaches the core of the cluster potential well. This trend is
consistent with that seen in other environmental studies, both in
the dependence of current star formation rate 
\citep[e.g.,][]{balogh98,kauf2004}, and the variation in the incidence of
recent star formation
\citep{nolan06}, on local galaxy density.  
However, the unusual feature, in
the case of galaxies belonging to inter-cluster filaments, is that on
top of this decline, there seems to be evidence for a sharp burst of
star formation at $\sim$3~$h_{70}^{-1}$ Mpc (1.5--2 times the virial
radius) from the centre of the nearest rich cluster.

It has been shown that galaxies can be surrounded by haloes of hot
X-ray emitting gas \citep[e.g.,][]{pederson2006}. However, it is
likely that many more galaxies have a warm halo of gas with
temperatures of $10^5\! - \! 10^6$~K, which is too cool to be detected
in X-ray observations.  This reservoir of warm gas would be expected to
provide for continued star formation in the galaxy, and its loss would
amount to the strangulation of star formation.

As a galaxy approaches the gravitational potential well of the rich
cluster at the extremity of a filaments, it would begin to encounter
the hot/warm intra-cluster Medium (ICM), possibly at a distance of the
order of the virial radius from its centre.  As the gaseous halo of
the galaxy interacts with the hot ICM, the galaxy would lose their
warm gas haloes through evaporation and possibly ram pressure
stripping.  This will lead to a steep decline in the SFR of galaxies
within $\sim 1.5-2 \,h_{70}^{-1}$~Mpc of the centre of the
cluster. This is seen in Fig.~\ref{peak-filament}.  In this region,
galaxy-galaxy harassment \citep{moore1999} will also be an
important effect. However, with most of the their reservoir of gas
already having been removed by this stage, star formation will not be
induced by the harassment, and the remaining gas of the galaxy will
continue to be stripped, leading to the observed steep dip in SFR in
the first two bins in Fig.~\ref{peak-filament} from the centre of the
cluster.

Even before this effect of the hot ICM begins to manifest itself on
the SFR of infalling galaxies, the density of galaxies will already
have begun to increase rapidly, which could be appreciable at
distances as large as twice the virial radius.  Approaching the outer
regions of the cluster along a filament, galaxies would experience
close interactions even before they experience any significant influence
of the cluster gravity and ICM.  Thus, the most likely cause for this
observed sudden burst in SFR at $\sim 3 \, h_{70}^{-1}$~Mpc from the
centre of the cluster, would be galaxy-galaxy harassment, which is a
rapidly acting process which works efficiently in crowded
environments. This effect would be found superposed on the general
trend of decrease in SFR towards the core of the cluster, as found in
galaxies elsewhere.  These galaxies are yet to encounter the hot and
dense ICM of the cluster, and thus would have not had their gas
stripped or evaporated, but the close interaction with other galaxies
would lead to density fluctuations in the gas, resulting in bursts of
star formation.  \citet{moore1999} show that harassment is a rapid
effect, which would account for the sharpness of the peak in the SFR
in this region. The range of distance over which this peak is seen is
a narrow one, also because its existence depends on both the existence
of substantial fuel for star formation, as well as sufficient external
influence which would act as trigger, and this occurs over a limited
range between the sparser expanse of the filament and the dense ICM of
the core of the cluster.

A photometric study \citep{haines06mn}
of the nearby Shapley supercluster
\citep{shap89} reveals an excess of faint
($M_R\!>\! -18$), bluer star-forming galaxies
at $\sim 1.5\, h_{70}^{-1}$~Mpc from the centres
of the rich clusters in the core of the supercluster.
At higher redshift,
\citet{moran2005} 
observe a similar sharp peak in SFR (evident from redshifted
[OII] emission) in the outskirts of the rich cluster CL0024 at $z\!=\!
0.4$, at $1.8\, h_{70}^{-1}$~Mpc from the cluster centre, where galaxy
harassment is cited as a possible cause.  While it is encouraging to
observe similar effects in other studies using different observables,
the SFR peak found in the case of the Shapley supercluster or  
CL0024 is a factor of 1.5--2 closer to the cluster
core than the distance at which the SFR peak is found in our study.
It is possible that the galaxies in the Pisces-Cetus filaments
experience the effect of galaxy-galaxy harassment at a larger distance
from the cluster centre, since the accretion of galaxies in this case
occurs along relatively narrow filaments, resulting in similar galaxy
densities further out than would be seen in the case of clusters where
galaxies are being accreted from all directions.

Fig.~\ref{sfveldis} reveals no discernible dependence of the position
of the peak in SFR on the velocity dispersion of the cluster involved,
which would have been expected if the SFR trigger had been dominated
by the gravitational effect of the cluster. Even if the tidal effects
of the cluster potential contributes to the enhanced SFR, it is
evidently dominated by local effects as described above.

The effect of ``backsplash'' \citep{gill2005}, which results from a
galaxy on an oscillating orbit at the bottom of the potential well at
the core of a cluster, is expected to be strongest at about
$\sim$1~Mpc from the centres of clusters, and is thus not likely to
have a strong effect on the peak in SFR seen in our filament galaxies,
which occurs further out in distance. However, it may mean that the
increase in SFR in some galaxies in the sharp peak  may be even
stronger than our mean suggests.

The observed peak in SFR may even contribute to our steep gradient
within $\sim3 \, h_{70}^{-1}$~Mpc, making it steeper.  The galaxies
that are star forming will have had their gas temperature raised and
thus when they encounter the ICM they will be more prone to
evaporation and stripping and hence have a rapidly decreasing SFR
\citep[for a similar effect seen in starburst galaxies in
a poor group, see ][]{rasmussen2006}.

The caveat in the use of the $\eta$ parameter as a proxy for current
SFR might be that the principal component analysis used in obtaining
the $\eta$ parameter might not have removed contributions from active
galactic nuclei (AGN). It therefore remains a possibility that part of
enhancement in mean $\eta$ is due to AGN activity rather than
conventional star formation, consistent with the results of
\citet{ruderman2005} who find a prominent spike in the number of
X-ray AGN in clusters at $\sim 2.5 \, h_{70}^{-1}$~Mpc. However, not
all X-ray detected AGN have the usual emission lines in their optical
spectra \citep{shen2007}.

Fig.~\ref{fil-mags} shows that the enhanced SFR is almost exclusively
seen in galaxies of lower luminosity (dwarfs) than in the giant
galaxies, as is indeed also seen in the study of CL0024 galaxies
\citep{moran2005}. Other studies of low-redshift galaxies
\citep[\eg][]{haines06apj} have also
revealed differences between the SFR properties of giant and dwarf
galaxies.  This supports our favoured scenario, since dwarf galaxies
are expected to be more vulnerable to galaxy harassment than giants.
\citet{moran2005} shows this, where almost all the galaxies in their peak in the [OII]
equivalent width are dwarfs (also see \citet{sato2006}). 
It also shows that the merging of
galaxies is unlikely to be a major contributor, in which case the SFR
would have been more enhanced in giants, which have a higher
cross-section for mergers.

\section{Conclusions}
\label{sec:conc}

We have explored the environmental dependence of star formation in
galaxies, belonging to supercluster filaments connecting rich clusters
of galaxies, in the Pisces-Cetus supercluster, which lies within the
2dF galaxy redshift survey. We have used the $\eta$ parameter, a
quantity derived from a principal component analysis
\citep{madgwick2002} of the 2dFGRS galaxy spectra, which is\ 
known to correlate well with the equivalent width of the H$\alpha$
emission line, as an index of current star formation rate within a
galaxy.

For galaxies belonging to three Pisces-Cetus filaments (each over 20
$h_{70}^{-1}$~Mpc long), connecting three pairs of rich clusters, we
have studied  the variation of the mean $\eta$ parameter with
distance along the filaments, from the cores of the rich clusters out to
the sparser reaches of the filament. We have also investigated
the variation of the fraction of passive galaxies with distance
from the cluster cores.

It is well known that the star formation rate in galaxies is the
lowest in the cores of rich clusters, and we confirm this by showing
that the value of mean $\eta$ falls to its lowest value, corresponding
to a maximum in the fraction of passive galaxies, for both giant and
dwarf galaxies in the cores of clusters, both in the supercluster
filaments, as well as elsewhere in the 2dFGRS.  Away from the cores of
clusters, while the SFR increases steadily as the local galaxy density
drops, everywhere in the 2dFGRS, there is an increased activity of
star formation, in a short distance interval between 3--4
$h_{70}^{-1}$~Mpc from the centre of the clusters, for the galaxies
belonging to the three Pisces-Cetus inter-cluster filaments studied
here.  This peak in star formation in filament galaxies is seen to be
almost entirely due to dwarf galaxies ($-20\!<\! M_B \!\le\! -17.5$).
The position of the peak does not depend on the richness of the
cluster.

We conclude that the most likely cause for this sharp enhancement in
SFR in the outskirts of the clusters is galaxy-galaxy harassment. This
occurs when galaxies falling into the cluster along narrow filaments
reach a certain density where close interactions between galaxies
become important.  For this to be effective, this has to occur before
the galaxy loses most of its gas to stripping and evaporation due to
the dense of ICM of the rich cluster. The predominance of SFR in dwarfs
indicates that galaxy merging is not an important factor.

Encouraged by the evidence of enhanced galaxy interaction found in the
luminosity function of groups \citep{miles04,miles06}, we have
examined the influence of the group environment on the variation of
SFR in our sample of galaxies. However, we find that while the group
members show the same general trend of gradually declining star
formation in denser environments, as seen in all galaxies in the
2dFGRS, the sharp enhancement in SFR in the outskirts of clusters, as
seen in this work, occurs at the same position for group members as
for non-group members on the Pisces-Cetus filaments, and is not seen
elsewhere.  This suggests that the enhancement in SFR is mostly
related to properties of the filament environment than the group
environment for the galaxies concerned.

Finally, although the results in this work are indicative of
interesting environmental effects on star formation in galaxies, where
the environment is characterised by parameters beyond the usual
projected local densities that are commonly found in the literature,
they come from a small sample of about a thousand galaxies in three
filaments in a single supercluster.  Moreover, we have used an
indirect measure of star formation (the $\eta$ parameter) derived from
the optical spectra of galaxies, and no allowance has been made for
the possible contribution of AGN to the spectral parameters. We look
forward to exploring these effects in a larger ensemble of filaments
from both the 2dFGRS and SDSS surveys, not only using various measures
of star formation, but also scrutinising the effects of a wider range
of environmental parameters that large samples would be able to
afford.


\section*{Acknowledgements}

We are indebted to the anonymous referee whose valuable suggestions
have led to a significant improvement of this paper.  We thank
Arif Babul and Trevor Ponman for very useful discussions,
Suketu Bhavsar for developing  the formalism used to find
the supercluster members, and those involved in creating the 2dFGRS
survey and the 2PIGG catalogues, which are the main sources of the
data used in this work.  Finally, we would like to thank all those
involved with the continual maintenance and enrichment of the
NASA/IPAC Extragalactic Database (NED), which is operated by the Jet
Propulsion Laboratory, California Institute of Technology, under
contract with the National Aeronautics and Space Administration.

\label{lastpage}

\clearpage

\end{document}